\documentclass[aps,prl,reprint,superscriptaddress]{revtex4-2}
\usepackage{graphicx,wrapfig}
\usepackage{fancyhdr}
\usepackage[colorlinks,citecolor = blue,linkcolor=blue]{hyperref}
\usepackage{ifthen}
\usepackage{color}
\usepackage{epsfig}
\usepackage{dcolumn}
\usepackage{float}
\usepackage{mathtools}
\usepackage{hyperref}
\usepackage{comment}
\usepackage{bbm}
\usepackage{amsmath}
\usepackage{soul}
\usepackage{ulem}

\newcommand{\TNSe}{{Ta$_2$NiSe$_5$}}
\newcommand{\TNS}{{Ta$_2$NiS$_5$}}
\newcommand{\TNSS}{{Ta$_2$Ni(Se,S)$_5$}}

\begin{document}


\title{Strong long-wavelength electron-phonon coupling in \TNSS}



\author{Zhibo Kang}
\affiliation{Department of Applied Physics, Yale University, New Haven, Connecticut 06511, USA}

\author{Burak Gurlek}
\affiliation{Max Planck Institute for the Structure and Dynamics of Matter and Center for Free-Electron Laser Science, Luruper Chaussee 149, 22761, Hamburg, Germany} 

\author{Weichen Tang} 
\affiliation{Physics Department, University of California, Berkeley, California 94720, USA}
\affiliation{Materials Sciences Division, Lawrence Berkeley National Lab, Berkeley, California 94720, USA}

\author{Xiang Chen}
\affiliation{Physics Department, University of California, Berkeley, California 94720, USA}
\affiliation{Materials Sciences Division, Lawrence Berkeley National Lab, Berkeley, California 94720, USA}

\author{Jacob P.C. Ruff}
\affiliation{Cornell High Energy Synchrotron Source, Cornell University, Ithaca, New York 14853, USA}

\author{Ahmet Alatas} 
\affiliation{Argonne National Laboratory, Argonne, Illinois 60439, USA}

\author{Ayman Said} 
\affiliation{Argonne National Laboratory, Argonne, Illinois 60439, USA}

\author{Robert J. Birgeneau}
\affiliation{Physics Department, University of California, Berkeley, California 94720, USA}
\affiliation{Materials Sciences Division, Lawrence Berkeley National Lab, Berkeley, California 94720, USA}
\affiliation{Department of Materials Science and Engineering, University of California, Berkeley, California 94720, USA}

\author{Steven G. Louie} 
\affiliation{Physics Department, University of California, Berkeley, California 94720, USA}
\affiliation{Materials Sciences Division, Lawrence Berkeley National Lab, Berkeley, California 94720, USA}

\author{Angel Rubio}
\affiliation{Max Planck Institute for the Structure and Dynamics of Matter and Center for Free-Electron Laser Science, Luruper Chaussee 149, 22761, Hamburg, Germany}
\affiliation{Initiative for Computational Catalysis (ICC), Flatiron Institute, New York, New York 10010, USA}

\author{Simone Latini}
\affiliation{Max Planck Institute for the Structure and Dynamics of Matter and Center for Free-Electron Laser Science, Luruper Chaussee 149, 22761, Hamburg, Germany}
\affiliation{Department of Physics, Technical University of Denmark, 2800 Kgs. Lyngby, Denmark}

\author{Yu He}
\email{yu.he@yale.edu}
\affiliation{Department of Applied Physics, Yale University, New Haven, Connecticut 06511, USA}


\date{\today}

\begin{abstract}
The search for intrinsic excitonic insulators (EI) has long been confounded by coexisting electron-phonon coupling in bulk materials. Although the ground state of an EI may be difficult to differentiate from density-wave orders or other structural instabilities, excited states offer distinctive signatures. One way to provide clarity is to directly inspect the phonon spectral function for long wavelength broadening due to phonon interaction with the high velocity EI phason. Here, we report that the quasi-one-dimensional (quasi-1D) EI candidate \TNSe~shows extremely anisotropic phonon broadening and softening in the semimetallic normal state. In contrast, such a behavior is completely absent in the broken symmetry state of \TNSe~and in the isostructural \TNS, where the latter has a fully gapped normal state. By contrasting the expected phonon lifetimes in the BCS and BEC limits of a putative EI, our results suggest that the phase transition in \TNSS~family is closely related to strong interband electron-phonon coupling. We experimentally determine the dimensionless coupling $\frac{g}{\omega_0}\sim10$, revealing \TNSS\ as a rare ``ultra-strong coupling'' material.
\end{abstract}


\maketitle

Excitonic insulators are materials in which electrons and holes spontaneously bound into excitons and subsequently form a thermodynamic condensate~\cite{Blatt1962BECofExcitons,Jerome1967Excitonic,Halperin1968}. So far, the closest realizations are in artificially engineered quantum Hall bilayer systems, which can exhibit perfect Coulomb drag in counterflow transport~\cite{Kellogg2004,nguyen2023}, anomalous zero-bias conductance peak in interlayer tunneling~\cite{Spielman2000}, and incompressible charge with compressible excitons~\cite{Ma2021,Gu2022}. However, the low-energy scale associated with such interlayer excitons often requires sub-Kelvin temperatures to reach condensation. Bulk crystalline materials could in principle host excitons with stronger binding energy thanks to smaller lattice site spacing \cite{Kazimierczuk2014-un,Hill2015,Vaquero2020-kd}, but interband electron-phonon coupling can create similar electron-hole excitations, destabilize the atomic lattice, and lead to a broken symmetry state with a charge redistribution indistinguishable from a bona fide excitonic insulator~\cite{Jerome1967Excitonic,Halperin1968}. The main hurdle of identifying a bulk excitonic insulator is to ascertain the causal roles of direct electron-hole Coulomb attraction and/or electron-phonon coupling behind the charge redistribution in the ground state~\cite{Kogar2017,Rossnagel2002Charge,Hedayat2019Excitonic,Wegner2020Evidence, Pashov2025TiSe2,Baldini2023, UEDTNSe2025}. This is made possible via independent controls of electron/hole distribution and population in engineered quantum Hall bilayer systems. In contrast, separately controlling the charges and comparing contributions from different degrees of freedom are more challenging in bulk materials. 

\begin{figure}[!tbh]
\includegraphics{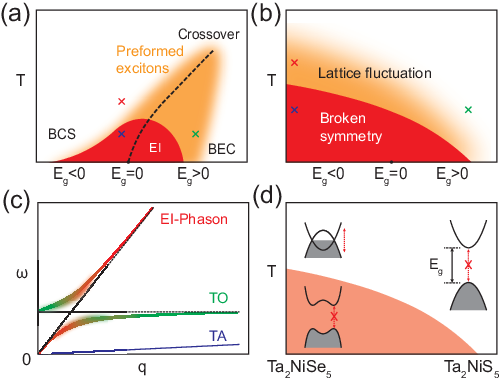}
\caption{\label{fig_schematics}\textbf{Phase diagrams of an ideal EI and \TNSS.} \textbf{(a)} The phase diagram of an ideal EI with a BCS-BEC crossover \cite{Jerome1967Excitonic,Seki2014ARPES,Lu2017NatureComm,Sugimoto2018Strong,Chen2020Doping}. $E_g$ is the band gap. \textbf{(b)} The phase diagram of \TNSS reported recently in \cite{Chen2023NatureComm}. The crosses indicate the measurement locations in this work. \textbf{(c)} Small $\mathbf{q}$ anticrossing between the massless phase mode (red, seen through charge susceptibility) and an optical phonon (green, seen through dynamic structure factor) \cite{Murakami2020Collective}. \textbf{(d)} Schematics of the low-energy band structure in \TNSS\ corresponding to the measurement positions denoted in \textbf{(b)}~\cite{Chen2023NatureComm}.}
\end{figure}

\begin{figure*}[!tbh]
\includegraphics{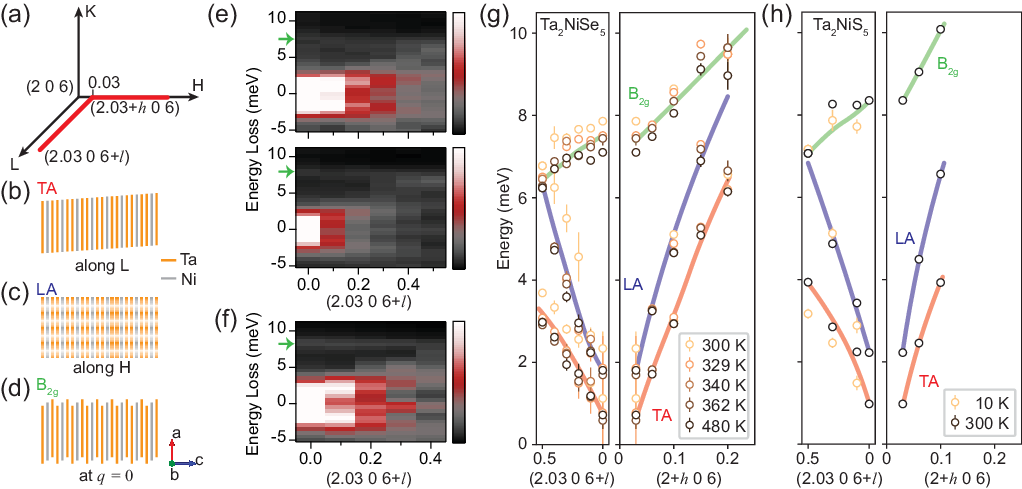}
\caption{\label{fig2}\textbf{Phonon dispersions in \TNSe~and \TNS.} \textbf{(a)} Schematics of the scan directions in reciprocal space. Schematics of the motions of the Ta (orange) and Ni (gray) chains of \textbf{(b)} the transverse acoustic (TA) and \textbf{(c)} the longitudinal acoustic (LA) phonon near $\textbf{q}=$0, and \textbf{(d)} the 2 THz $\text{B}_\text{2g}$ phonon at $q=0$. \textbf{(e-f)} The energy-loss spectrum maps of \TNSe~at 362 K (\textbf{(e)} top) and 300 K (\textbf{(e)} bottom) and \TNS~at 300 K \textbf{(f)} from the IXS along the momentum trajectory (2.03 0 6+$l$). The green arrows indicate the location of the 2 THz $\text{B}_\text{2g}$ phonon. \textbf{(g)} Phonon dispersion for \TNSe~at different temperatures across the phase transition. \textbf{(h)} Phonon dispersion for \TNS~at 10 K and 300 K. The smooth lines are guides to the eye.}
\end{figure*}

Among the latest excitonic insulator candidates is the quasi-1D ternary chalcogenide \TNSe, which is believed to also host strong electron-phonon coupling (EPC)\cite{Baldini2023, UEDTNSe2025,Pavel2021,ChenPRR2023,Chen2023NatureComm,yan2019strong}. At $T_\text{S}=329$ K, \TNSe\ undergoes a second-order orthorhombic-to-monoclinic transition, concurrent with a band gap opening of at least 150 meV \cite{diSalvo1986physical,PBEDFTSimone2021,Baldini2023,Pavel2021,ChenPRR2023, Lu2017NatureComm, Larkin2017Giant, He2021Tunneling}. Below $T_\text{S}$, the flat valence band top in the ground state was initially taken as evidence for exciton condensation \cite{Wakisaka2009ARPES,Seki2014ARPES}, yet direct electron–hole Coulomb attraction was later shown to be too weak to account for the gap \cite{ChenPRR2023}. Above $T_\text{S}$, although it was reported that the normal state is a narrow-gap semiconductor \cite{Seki2014ARPES,Fukutani2021Detecting, Wakisaka2012Photoemission}, more recent investigations gravitate towards a semimetallic state with a spectral pseudogap \cite{Watson2020PRR, Tang2020Non, Li2024Disentangling} which is ascribed to strong lattice fluctuations \cite{ChenPRR2023,Chen2023NatureComm,Baldini2023} or preformed excitons \cite{kim2021direct}. No phonon softening is seen at $2k_F$~\cite{Nakano2018,Pavel2021,kim2021direct}, contradicting expectations in an ideal Peierls transition \cite{peierls1955quantum,ZhuPNAS2015}. Meanwhile, a transverse acoustic phonon softens at $\mathbf{q}\rightarrow0$ along $c^*$~\cite{Nakano2018} -- similar to that in a cooperative Jahn-Teller transition~\cite{Birgeneau1970,NeutroninCeCu6} -- but is contended to enhance rather than drive the transition \cite{Pavel2021}. On the other hand, a 2-THz $\text{B}_\text{2g}$ optical phonon develops strong Fano shape near and above $T_\text{S}$, indicating strong coupling to an unspecified excitation continuum \cite{Pavel2021,kim2021direct}. Thus, directly examining the phonon spectral function to elucidate the electronic coupling - and any potential excitonic coupling - to low-energy phonons is essential not only for resolving the excitonic-insulator debate, but also for assessing whether the coupling is sufficient to invalidate the Born–Oppenheimer approximation.

The low-energy phonon dynamic structure factor in momentum space carries distinct fingerprints of coupling to single particle excitations, preformed excitons, or the phase excitation -- phason -- in an EI \cite{Ye2021Raman,kim2021direct,Pavel2021,Kaneko_ANewEra}. By comparing the ideal excitonic insulator phase diagram \cite{Jerome1967Excitonic,Seki2014ARPES,Lu2017NatureComm,Sugimoto2018Strong,Chen2020Doping} with the strong electron-phonon coupling phase diagram reported recently \cite{Chen2023NatureComm}, the differences can be further illustrated. Two outcomes are expected in a bona fide EI through phason–phonon coupling: (i) small-$\mathbf{q}$ hybridization when electron–phonon coupling is weak, and (ii) global phonon softening when the coupling is strong \cite{Murakami2020Collective,Kaneko_ANewEra} (Fig.~\ref{fig_schematics}(a,c)). In contrast, for an interband electron-phonon coupling scenario, the low-energy phonon is expected to become increasingly long-lived as the single-particle band gap increases. Tuning the normal-state single-particle gap from negative to positive should therefore differentiate interband electron–phonon coupling from exciton–phonon coupling effects through low-energy dynamic structure factor. In \TNSS, increasing S content drives the normal state from a $\sim$300 meV negative-gap semimetal to a $\sim$300 meV positive-gap semiconductor \cite{Lu2017NatureComm,Chen2023NatureComm}, while $T_S$ is suppressed to below 10 K upon full S substitution (Fig.~\ref{fig_schematics}(b,d)) \cite{diSalvo1986physical,Chen2023NatureComm}. Accordingly, if truly an EI, low-energy phonons should broaden most in the broken-symmetry state of \TNSe\ due to anticrossing with the steeply dispersing phason (Fig.~\ref{fig_schematics}(a), blue cross). If the phonon broadening is mainly caused by coupling to preformed excitons, it would then be most pronounced in the fluctuating state towards the BEC limit (Fig.~\ref{fig_schematics}(a), green cross). Conversely, if interband electron–phonon coupling dominates the transition, phonons should broaden most in the normal state of \TNSe\ (Fig.~\ref{fig_schematics}(b), red cross) but becomes long-lived in the ground state of \TNSe\ or in \TNS\ (Fig.~\ref{fig_schematics}(b), blue/green crosses). In this work, we test the role of the EI correlations in the phase diagram of \TNSS\ based on this criterion.

Our x-ray scattering results show highly anisotropic diffuse signal between (2 0 6) and (2 0 8) Bragg peaks near $T_S$, which could come from the softening and/or the lifetime reduction of low-energy phonons (for the discussion to rule out static disorder effect, see Appendix). Moreover, a recent ultrafast electron diffraction (UED) experiment reported strong temporal fluctuation of the (2 0 6) and (2 0 8) peak intensities \cite{UEDTNSe2025}. To clarify whether this is due to phason-phonon coupling in an EI, we investigate the dynamic structure factor along (2.03 0 6+$l$) and (2+$h$ 0 6) directions (Fig.~\ref{fig2}(a), see Supplementary Note I for the lattice structure and the definition of the Miller indices \cite{supp}) in both \TNSe~and \TNS~with inelastic x-ray scattering (IXS). Two leading phonon candidates are the 2 THz $\text{B}_\text{2g}$ optical mode and the transverse acoustic mode~\cite{Pavel2021,Baldini2023, UEDTNSe2025,PBEDFTSimone2021}, both involving shearing of the Ta chains (Fig.~\ref{fig2}(b-d), see also Supplementary Note VIII \cite{supp}). Figure~\ref{fig2}(e-f) show the x-ray energy-loss spectra of \TNSe~and \TNS~ along (2.03 0 6+$l$). Four low-energy phonon modes can be identified (see Supplementary Note II for damped harmonic oscillator (DHO) fitting \cite{supp}): the transverse acoustic (TA) and longitudinal acoustic (LA) modes, the 2-THz $\text{B}_\text{2g}$ and the 3-THz $\text{B}_\text{2g}$ optical modes \cite{Ye2021Raman,Pavel2021,kim2021direct,Nakano2018}. No dramatic phonon softening is observed in \TNSe~across the phase transition (Fig.~\ref{fig2}(g), see also Supplementary Note III for detailed temperature dependence \cite{supp}). Compared to the extensive acoustic phonon softening along (4 0 $l$) \cite{Nakano2018}, the softening disappears almost completely after only 0.03 r.l.u. offset along H (see Supplementary Notes IV for temperature dependent sound velocities and VII for schematics of such an anisotropic softening \cite{supp}). This indicates that the lattice fluctuation associated with the phase transition involves shear motion of rigid Ta chains of at least 30 unit cells long.

\begin{figure}
\includegraphics{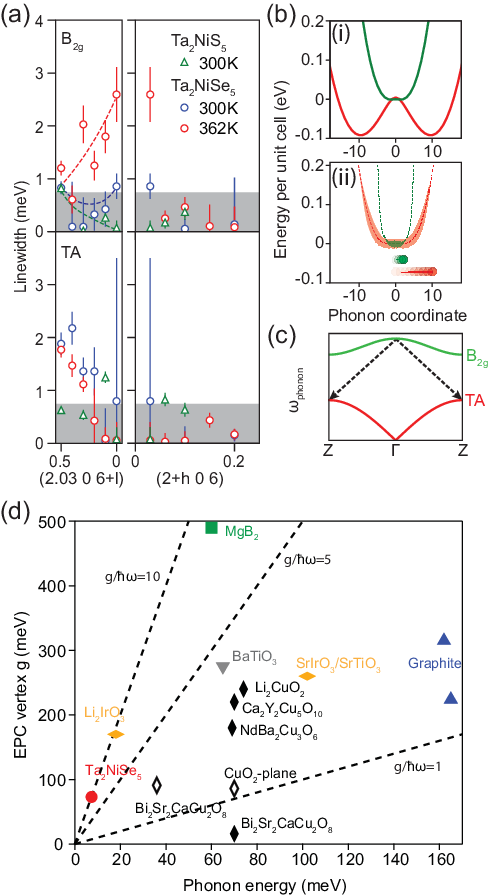}
\caption{\label{fig4}\textbf{Highly anisotropic phonon linewidths in \TNSe~and \TNS.} \textbf{(a)} The phonon linewidths of $\text{B}_\text{2g}$ (top) and TA (bottom) modes fitted from the energy-loss spectra. The dashed lines are guides to the eye. The gray shade indicates the half width at half maximum of the energy resolution function. \textbf{(b)} Ab initio calculation of phonon-coordinate-dependent free energy potentials for the $\text{B}_\text{2g}$ mode in \TNSe~(red) and \TNS~(green) (i) with and (ii) without the harmonic component. Weights are schematic depictions of physically accessed phonon coordinates in real systems.
\textbf{(c)} Schematic decay channel for the $\mathbf{q}=0$ $\text{B}_\text{2g}$ phonon. \textbf{(d)} Comparing the largest electron-phonon coupling vertex g in phonon momentum space and the phonon energy in different material systems (hollow black rhombi: cuprate calculations \cite{Johnston2010Systematic,Devereaux2004Anisotropic}, solid black rhombi: cuprates \cite{Qin_EPCdetermining_2010,Rossi2019Experimental,Braicovich2020Determining,Lee2013Role,Lee2014Charge,Johnston2016Electron}, orange rhombi: iridates \cite{Meyers2018Decoupling,Vale2019High}, blue triangles: graphites \cite{Piscanec2004Kohn,Na2019Direct}, grey triangle: BaTiO$_3$ \cite{Fatale2016Hybridization}, green square: MgB$_2$ \cite{Shukla2003Phonon}), among which \TNSe~shows the largest ratio $\sim 10$}.
\end{figure}

We turn to the optical phonons next. As shown in Fig.~\ref{fig2}(g), no obvious 2 THz $\text{B}_\text{2g}$ phonon softening is observed in \TNSe~at any momentum across the phase transition. Moreover, in the isostructural \TNS, aside from an overall hardening of all phonon energies due to the heavy Se being replaced by the lighter S, no qualitative difference is seen in the low-energy phonon dispersions (Fig.~\ref{fig2}(h)).

However, the phonon linewidths exhibit an anomaly at the long wavelength limit. Figure~\ref{fig4}(a) shows the phonon linewidths of the TA and 2-THz $\text{B}_\text{2g}$ modes after deconvolving the 1.5~meV instrument resolution \cite{Said30ID,Toellner30ID}. In the normal state of \TNSe, the linewidth of the 2 THz mode increases substantially at $\mathbf{q}\rightarrow0$ along L while remaining resolution-limited along H. In contrast, it is always resolution-limited in the broken symmetry phase of \TNSe~and in \TNS. 
Clearly, the 2~THz mode is uniquely short-lived only in the normal state of \TNSe, where strong lattice fluctuations and a large electronic pseudogap have been reported, and the interband electron-phonon coupling is strongest \cite{ChenPRR2023,Chen2023NatureComm}.

An increase of phonon linewidth implies strong scattering of the phonons, which can come dynamic processes such as phonon-phonon interaction~\cite{May1998Anharmonic} and electron-phonon interaction~\cite{grimvall1981electron}. In the specific case of an excitonic insulator, optical phonons can also interact with the phason of the exciton condensate \cite{Murakami2020Collective} (Fig.~\ref{fig_schematics}(c)). Our density-functional calculation shows in \TNS~a double-well potential, along the $\text{B}_\text{2g}$ mode, so shallow ($\sim 1$ meV) that no structural phase transition would occur down to the lowest experimental temperature reached in previous studies (Fig.~\ref{fig4}(b)(i))~\cite{Chen2023NatureComm,Lu2017NatureComm,Ye2021Raman}. Meanwhile, the computed anharmonic component of the potential for the 2 THz phonon is $\sim 150$ times larger in \TNSe~than in \TNS~at their respective broken symmetry positions (Fig.~\ref{fig4}(b)(ii)). 
This in part reflects the much larger phonon fluctuation $\langle Q^2\rangle$ in \TNSe~than in \TNS~\cite{Chen2023NatureComm}, despite similar oscillator masses and frequencies (Fig.~\ref{fig2}). Notably, this calculation does not include any excitonic effect and already captures major experimental observations (see Supplementary Note IX \cite{supp}). Dynamic phonon-phonon interaction could in principle also cause phonon lifetime reduction. For example, the zone center 2 THz phonon could resonantly decay into TA phonons at Z points, schematically shown in Fig.~\ref{fig4}(c). This decay channel is allowed by energy-momentum conservation in both \TNSe~and \TNS~according to the measured phonon dispersions in Fig.~\ref{fig2}. But the $\mathbf{q}\rightarrow0$ phonon broadening is only present in \TNSe. On the other hand, the coupling of phonons to preformed excitons is also suggested in a previous study~\cite{kim2021direct}. However, strongest excitonic fluctuation is expected in the normal state of \TNS~-- the supposed BEC limit. This is opposite to our observation. Finally, the interaction between the EI phason and optical phonons shall lead to broadened phonon linewidths at $\mathbf{q}\sim\frac{\omega_0}{\hbar v_\mathrm{phason}}\sim0.03 c^*$, along with a decreased phonon energy~\cite{Murakami2020Collective}. This manifests as $\mathbf{q}\rightarrow0$ broadening with IXS in the broken symmetry state (Fig.~\ref{fig_schematics}(c)), since the phason velocity $v_\text{phason}$ would be dictated by electronic energy scales ($\sim10^5$m/s) compared to the speed of sound ($\sim10^3$m/s). In contrast, the dramatic decrease of the $\text{B}_\text{2g}$ 2 THz phonon linewidth (Fig.~\ref{fig4}(a)) and the increase of the phonon energy (Fig.~\ref{fig2}(g)) in the broken symmetry state of \TNSe~deem phonon-phason coupling unlikely. These observations, in addition to the experimentally measured \TNSS~phase diagram depicted in Fig.~\ref{fig_schematics}(d)~\cite{Chen2023NatureComm}, prompt strong interband electron-phonon interaction as a serious consideration behind the phase transition, as inferred in previous Raman \cite{Pavel2021}, electron diffraction \cite{UEDTNSe2025} and angle-resolved photoemission spectroscopy (ARPES) studies \cite{Baldini2023,ChenPRR2023,Chen2023NatureComm}. 
Here, the semimetallic normal state of \TNSe~offers considerable low-energy electronic states for such scattering to occur, which naturally results in a large phonon linewidth. While in the broken symmetry state of \TNSe~and the semiconducting normal state of \TNS, such low-energy excitations are strongly suppressed by the $\sim300$ meV energy gap, where long phonon lifetime is restored.

Furthermore, based on the assumption that most of the phonon linewidth primarily comes from the interband electron-phonon coupling, we may now provide a direct estimate of the electron-phonon coupling vertex $g_{\mathbf{q}\rightarrow0}$ by combining the phonon linewidth measured by IXS with the single-particle band structure fitted from ARPES~\cite{Chen2023NatureComm, Pavarini:837488}. Within linear response~\cite{Allen1972,Qin_EPCdetermining_2010}, the momentum-dependent phonon broadening $\Gamma(\textbf{q})$ induced by EPC can be written as \cite{Albers1976EPC, Giustino2017EPC}:
\begin{eqnarray}\label{eq1}
\Gamma(\textbf{q})=&&4\pi\sum_{mn}\int\frac{d\textbf{k}}{\Omega_{BZ}}\vert g_{mn}(\textbf{k},\textbf{q})\vert ^2 (f_{n\textbf{k}}-f_{m\textbf{k}+\textbf{q}})\nonumber\\
&&\times
\delta(\epsilon_{m\textbf{k}+\textbf{q}}-\epsilon_{n\textbf{k}}-\hbar\omega_{\textbf{q}}),
\end{eqnarray}
where $\Omega_{BZ}$ is the volume of the Brillouin zone, $f_{n\textbf{k}}=f(\epsilon_{n\textbf{k}})$ is the Fermi function, $\epsilon_{n\textbf{k}}$ is the electron dispersion and $\hbar\omega_{\textbf{q}}$ is the phonon energy at $\textbf{q}$.

\begin{table}[H]
\caption{\label{tab:table1}%
Summary of the phonon energy $\hbar\omega_{\textbf{q}_0}$, the phonon linewidth $\Gamma(\textbf{q}_0)$; the imaginary part of the Lindhard function Im$\chi(\textbf{q}_0,\omega_{\textbf{q}_0})$, the corresponding calculated EPC vertex $\vert g(\textbf{q}_0)\vert$ at $\textbf{q}_0=0.03\times 2\pi/a$; the density of state at Fermi energy $N(0)$ and the (isotropic) EPC constant $\lambda$.
}
\begin{ruledtabular}
\begin{tabular}{ccc}
 & \textrm{\TNSe~}& \textrm{\TNS~} \\
\colrule
$\hbar\omega_{\textbf{q}_0}$ (meV) & 7.4$\pm$0.2 & 8.35$\pm$0.05 \\
$\Gamma(\textbf{q}_0)$ (meV) & 2.59$\pm$0.52 & 0.07$\pm$0.13 \\
Im$\chi(\textbf{q}_0,\omega_{\textbf{q}_0})$ (eV$^{-1}$)& 0.038 & 0.0004 \\
$\vert g(\textbf{q}_0)\vert$ (meV) & 73$\pm$7 & 121$\pm$111 \\
N(0) (eV$^{-1}$)& 1.33 & 0\footnote{For semiconductor, N(0) is by definition 0.} \\
$\lambda$ & 0.97$\pm$0.17 & 0.26$\pm$0.01\footnote{For semiconductor, $\lambda$ is derived by comparing the DFT band mass with the ARPES measured band mass \cite{Chen2023NatureComm} based on methods proposed in \cite{Giustino2020EffectiveMass,Ponce2025EPhRenormalization}.} \\
\end{tabular}
\end{ruledtabular}
\end{table}

Assuming the EPC vertex $g_{mn}(\textbf{k},\textbf{q})=g(\textbf{q})$ is independent of the electron momentum $\textbf{k}$ and the band indices $m$, $n$ since only lowest energy conduction-valence bands are considered, Eqn.\ref{eq1} becomes
\begin{eqnarray}\label{eq2}
\Gamma(\textbf{q})
&& = 4\pi \vert g(\textbf{q})\vert^2 \textrm{Im}\chi(\textbf{q},\omega_{\textbf{q}}),
\end{eqnarray}
where $\chi(\textbf{q},\omega)$ is the Lindhard charge response function.
Here, we 
use experimentally fitted band dispersions $\epsilon_{m\mathbf{k}}$ \cite{ChenPRR2023,Chen2023NatureComm} to calculate $ \textrm{Im}\chi(\textbf{q},\omega_{\textbf{q}})$.
The EPC vertex and the dimensionless EPC constant are then estimated using the measured phonon energy and linewidth (Supplementary Note V and VI \cite{supp}). The results are summarized in Table \ref{tab:table1}. Notably, the experimentally derived vertex of 73 meV is in excellent agreement with previous first principles calculation~\cite{ChenPRR2023}. Given the exceptional ratio of $\frac{g}{\omega_{\mathbf{q}_0}}\sim10$, it is among the largest numbers in strong coupling/correlated material families (Fig. \ref{fig4}(d)) and falls into the ``ultra-strong'' coupling regime introduced in the cavity QED community~\cite{schlawin2022cavity}. This makes the \TNSS~family an exciting low dimensional solid-state platform to realize and investigate ultra-strong coupling phenomena \cite{schlawin2022cavity}, such as squeezed phonon states \cite{michael2024photonic,Haque2024Terahertz}, phonon entanglement \cite{Liu2025Entanglement}, polaronics engineering \cite{Zhang2023Bipolaronic} and other beyond-Born-Oppenheimer-approximation behaviors~\cite{wang2021fluctuating,ChenPRR2023}. In such a regime, the lattice is no longer a spectator of electronic processes, but rather an extremely effective control knob, and long-wavelength static Coulomb screening becomes ineffective. This may also offer a natural explanation to observations of carrier dynamics faster than the 2 THz phonon period, since in the strong coupling regime, the time scale is set by the much larger coupling vertex $g$. Excitingly, a series of very recent works predict and show pronounced long-wavelength strain-dependent low-energy electronic states in this system \cite{Shi2025Uniaxial,Tang2025Nonexcitonic}, further validating the physical consequence of this strong coupling beyond only a nominal classification scheme.

In conclusion, through IXS, we investigate the low-energy phonon dynamic structure factor at both ends of the phase diagram for \TNSS.
We reveal that the anisotropic phonon broadening and softening only happen in the semimetallic normal state of \TNSe, but not in the semiconducting broken symmetry state of \TNSe~or in the isostructural \TNS. This suggests the phase transition being closely associated with interband electron-phonon coupling rather than excitons or EI-phasons. In addition, through a direct experimental estimate of the EPC vertex for the 2-THz $\text{B}_\text{2g}$ shear phonon, we place the \TNSS\ family in the ``ultra-strong'' coupling regime introduced in cavity quantum materials \cite{ChenPRR2023,Chen2023NatureComm, Baldini2023,PBEDFTSimone2021}. 
Finally, our approach establishes high-resolution phonon spectroscopy as a general discriminator of intertwined phases that share similar ground-state properties through their distinct phonon–elementary excitation coupling.

\subsection*{Acknowledgments}
We thank Leonid Glazman, Lukas Windg{\"a}tter, Shuolong Yang, Fred Walker, Yu Song, Pavel Volkov, Zhenglu Li, Yao Wang and Yuta Murakami for helpful discussions. The work at Yale University was supported by National Science Foundation under grant DMR-2239171. This research used 30-ID HERIX beamline at the Advanced Photon Source, Argonne National Laboratory, supported by the U.S. Department of Energy under contract no. DE-AC02-06CH11357. Research conducted at the Center for High-Energy X-ray Science (CHEXS) was supported by the National Science Foundation (BIO, ENG and MPS Directorates) under award DMR-1829070. Work at Lawrence Berkeley National Laboratory was funded by the U.S. Department of Energy, Office of Science, Office of Basic Energy Sciences, Materials Sciences and Engineering Division under Contract No. DE-AC02-05-CH11231 within the Quantum Materials Program (KC2202) and within the Theory of Materials Program (KC2301) which provided the DFT calculations. The DFT frozen phonon calculations were performed using computation resources at the National Energy Research Scientific Computing Center (NERSC). We acknowledge support from Cluster of Excellence `CUI: Advanced Imaging of Matter'- EXC 2056 - project ID 390715994 and the Max Planck-New York City Center for Non-Equilibrium Quantum Phenomena. The Flatiron Institute is a division of the Simons Foundation. We acknowledge support from the European Union Marie Sklodowska-Curie Doctoral Network
SPARKLE grant No. 101169225. The authors also acknowledge the Brookhaven National Laboratory-Yale partner user agreement PU-313536.

\subsection*{Data Availability}
All data generated in this study have been deposited in the Figshare database with open access \cite{data}.

\section*{Appendix - distinguish static and dynamic effects behind phonon broadening}
\begin{figure}
\includegraphics{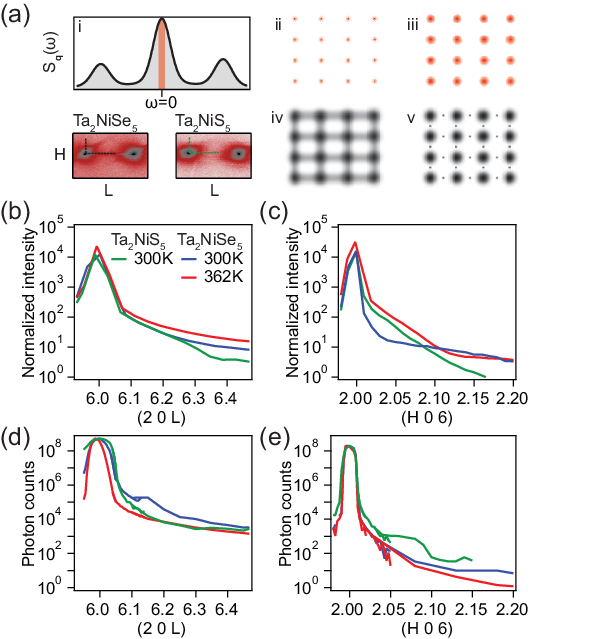}
\caption{\label{fig1}\textbf{Diffuse and quasi-elastic x-ray scattering in \TNSe~and \TNS.} \textbf{(a)} Schematics showing the differences between the quasi-elastic peak intensity and the diffuse signal (DS). \textbf{i} Upper panel: a typical energy-loss curve at non-Bragg condition with a quasi-elastic peak at $\omega=0$ and Stokes/anti-Stokes peaks of phonon excitation. The red shade indicates the quasi-elastic intensity. Lower panels: false color plots of XRD signal for \TNSe~and \TNS~around (2 0 6) and (2 0 8) peaks. The black and green dashed lines indicate the positions of the DS and quasi-elastic peak momentum cuts shown below. Note the unidirectional streak between Bragg peaks in \TNSe. \textbf{ii} and \textbf{iii} schematically shows Bragg peak broadening in the reciprocal space when isotropic static disorders are introduced into a perfect lattice. The gray shade shows the energy integrated intensity probed by x-ray diffraction (XRD). \textbf{iv} and \textbf{v} show typical patterns of DS and charge density wave (CDW) respectively. \textbf{(b-c)} DS in \TNSe~along  \textbf{(b)} (2 0 L) and \textbf{(c)} (H 0 6). Red, blue, and green lines are \TNSe~at 360 K, \TNSe~at 300 K and \TNS~at 295 K respectively. \textbf{(d-e)} Quasi-elastic peak intensity along \textbf{(d)} (2 0 L) and \textbf{(e)} (H 0 6).}
\end{figure}

We first use x-ray diffuse scattering (DS) to reveal any structural deviation from a perfect periodic lattice~\cite{James1948}. This deviation may be due to static impurities such as vacancies or dislocations, or dynamic atomic displacements such as phonons (Fig.~\ref{fig1}(a)). Figure~\ref{fig1}(b-c) show the DS intensity of \TNSe~above and below the transition temperature at 329 K and \TNS~at 300 K along H ($a^*$) and L ($c^*$) directions, respectively (for the lattice structure, see Supplementary Note I \cite{supp}). One general observation is that the DS along L is more pronounced than along H. Besides the anisotropy, the DS of \TNSe~above the transition temperature (red line) is stronger than below the transition (blue line) both along L (Fig.~\ref{fig1}(b)) and along H (Fig.~\ref{fig1}(c)). In contrast, the DS of \TNS~(green line) is much weaker than \TNSe~(blue line) along both directions away from the Bragg positions (Fig.~\ref{fig1}(b-c)).

While the DS is found to be stronger perpendicular to the chain, it is important to ascertain if this originates from phonons or from static effects such as impurities or dislocations. Taking advantage of the meV energy resolution in inelastic x-ray scattering, the static component can be extracted from the quasi-elastic channel \cite{He2025Resolving}. As shown in Fig.~\ref{fig1}(d-e), the quasi-elastic intensity is stronger along L than along H for both \TNSe~and \TNS. This can be understood through the quasi-1D nature of \TNSS, where dislocations between the chains (L direction) occur more frequently than along the chains (H direction). Expectedly, the quasi-elastic intensity increases upon cooling due to static dislocation freezing, which evolves oppositely to the total DS intensity (Fig.~\ref{fig1}(b-c)). This confirms that the DS is dynamic in nature, and it is contributed by phonons. Therefore, the narrow streak of diffuse intensity concentrated along L direction, which was also seen previously~\cite{Nakano2018,ChenPRR2023}, suggests highly anisotropic phonon anomalies. 

\bibliography{apssamp}

\end{document}